\documentclass[letterpaper,american,reprint, aps]{svjour3}
\usepackage[T1]{fontenc}
\usepackage[latin9]{inputenc}
\usepackage{url}
\usepackage{amsmath}
\usepackage{amssymb}
\usepackage[numbers]{natbib}

\makeatletter

\pdfpageheight\paperheight
\pdfpagewidth\paperwidth

\makeatother

\usepackage{babel}
\begin{document}
\title{Unitary toy qubit transport model for black hole evaporation}
\author{Bogus\l aw Broda}
\institute{Department of Theoretical Physics, Faculty of Physics and Applied
Informatics, University of \L ód\'{z}, 90-236 \L ód\'{z}, Pomorska
149/153, Poland; boguslaw.broda@uni.lodz.pl; http://merlin.phys.uni.lodz.pl/BBroda}
\date{5 December 2019}
\maketitle
\begin{abstract}
In a recent paper Osuga and Page have presented an explicitly unitary
toy qubit transport model for transferring information from a black
hole to the outgoing radiation. Following their idea we propose a
unitary toy model which involves (fermionic) Hawking states.
\end{abstract}

\section{Introduction}

The black hole information (loss) paradox/problem/puzzle concerns
difficulties in answering the question: ``Is black hole evaporation
unitary?'' Various answers and explanations have been proposed to
date, ``for'' its unitarity as well as ``against'' it (for recent
reviews see \citep{Harlow2016a,Marolf2017a,Polchinski2017c,Chakraborty2017a}).
(The latter possibility, i.e., loss of unitarity, presumably requires
some radical modifications of quantum mechanics, and presently this
option seems to be less popular, however see \citep{Unruh2017}).
To analyze the issue of unitarity in the context of black hole evaporation,
the idea of toy qubit  models has been proposed, developed and popularized,
especially in \citep{Giddings2012,Giddings2012a,Giddings2013d,Mathur2009,Mathur2009b},
and nicely reviewed in \citep{Avery2013}. Toy qubit  models give
a possibility to mimic, in a simplified way, black hole evolution,
and to trace, in detail, their (depending on the model) breaking or
maintaining unitarity.

In a recent paper \citep{Osuga2018}, Osuga and Page have proposed
a strikingly simple toy qubit model demonstrating transferring information
from a black hole to the outgoing radiation, which is explicitly unitary.
Characteristic features of their model, in particular, include: (\emph{i})
Hilbert spaces (implicitly) involved in their analysis are actually
fixed, though in their final state $\left|\Psi_{1}\right\rangle $
some part is dropped; (\emph{ii}) the model has a simple (tensor)
product structure (no mixing of different modes ``$i$''); and most
noteworthy, (\emph{iii}) unitarity of their model is shown in the
most direct way, i.e., a corresponding unitary operator is explicitly
constructed.

In the present paper, we propose another toy qubit transport model
for black hole evaporation, which is also explicitly unitary. Additional
nice characteristic features include: (\emph{i}) precisely (fermionic)
Hawking states are involved; (\emph{ii}) classical shrinking of a
black hole, in quantum formalism, corresponds to transition to a vacuum
state (in the black hole sector); (\emph{iii}) primary entanglement
(in the Hawking states) between modes inside and outside of the black
hole vanishes. 

For reader's convenience, as well as for ours, we follow conventions
and notation of \citep{Osuga2018} as closely as possible. In particular,
we denote black hole qubits (once the black hole forms) by $a_{i}$,
where $i=1,2,\ldots,n$, whereas qubits for the infalling radiation
and for the outgoing modes, by $b_{i}$ and $c_{i}$, respectively,
where $i=1,2,\ldots,N$. For technical simplicity, from now on we
confine ourselves to a Schwarzschild black hole, but this restriction
is not crucial for our further considerations.

\section{The model}

In the beginning, to each black hole mode $a_{i}$, we associate a
(fermionic) Hawking state $\left|H\left(\Omega_{i}\right)\right\rangle _{b_{i}c_{i}}$,
i.e., a pair of entangled radiation qubits, $b_{i}$ (infalling) and
$c_{i}$ (outgoing), in the state 
\begin{equation}
\left|H\left(\Omega_{i}\right)\right\rangle _{b_{i}c_{i}}=\cos\Omega_{i}\left|0\right\rangle _{b_{i}}\left|0\right\rangle _{c_{i}}+\sin\Omega_{i}\left|1\right\rangle _{b_{i}}\left|1\right\rangle _{c_{i}},\label{eq:hawking_state}
\end{equation}
which is precisely the fermionic state created by a black hole, according
to the Hawking effect. Alternatively, we could interpret (\ref{eq:hawking_state})
in terms of a two-term (qualitative) approximation of the bosonic
Hawking state (see, e.g., the second sentence after Eq.(2.2) in \citep{Mathur2011}).
The parameter $\varOmega_{i}$ is a function of Bogoljubov coefficients
following from space-time geometry. From more physical point of view,
$\varOmega_{i}$ is a $i$-mode dependent function of gravitational
field embodied (in the Schwarzschild case) by the black hole mass
$M_{\textrm{bh}}$.  The above association is equivalent to the assumption
that the black hole modes $a_{i}$ and the Hawking states (\ref{eq:hawking_state})
can be paired, i.e., $N=n$. It is true (or even obvious), at least
approximately, i.e.\ $N\sim n$. Indeed, the number of black hole
modes $n\sim S_{\textrm{BH}}=4\pi M_{\textrm{bh}}^{2}$, where $S_{\textrm{BH }}$
is the Bekenstein\textendash Hawking entropy (we have used Planck
units in which $\hbar=c=G=k_{\textrm{Boltzmann}}=1$), whereas estimated
number of outgoing radiation modes $N\sim M_{\textrm{bh}}^{2}$ (see
\citep{Mueck2016,Broda2017}). Consequently, the initial quantum state
of black hole modes and Hawking radiation is (cf.\ \citep{Osuga2018})
\begin{equation}
\left|\Psi_{0}\right\rangle =\sum_{q_{1},q_{2},\ldots,q_{n}=0}^{1}A_{q_{1}q_{2}\cdots q_{n}}\prod_{i=1}^{n}\left|q_{i}\right\rangle _{a_{i}}\prod_{i=1}^{n}\left|H\left(\Omega_{i}\right)\right\rangle _{b_{i}c_{i}},\label{eq:initial_state}
\end{equation}
where $A_{q_{1}q_{2}\cdots q_{n}}$ are the amplitudes for the black
hole modes $a_{i}$.

Let us note that the state $\left|\varPsi_{0}\right\rangle $ is a
tensor product of $n+1$ (block-)states, which are, in general, entangled
states. Schematically, we can express it explicitly as $\left|\varPsi_{0}\right\rangle =\left|A\right\rangle \otimes\left|H\right\rangle _{1}\otimes\cdots\otimes\left|H\right\rangle _{n}$.

The elementary, i.e., for fixed $i$, process we propose to consider
is (cf.\ Eq.(3.3) in \citep{Osuga2018}, and possibly also Eq.(3.3)
in \citep{Almheiri2013})
\begin{equation}
\left|q_{i}\right\rangle _{a_{i}}\left|H\left(\Omega_{i}\right)\right\rangle _{b_{i}c_{i}}\longmapsto\left|0\right\rangle _{a_{i}}\left|0\right\rangle _{b_{i}}\left|q_{i}\right\rangle _{c_{i}}.\label{eq:quantum_process}
\end{equation}
Obviously, the whole qubit transport process is a (``tensor'') product
of $n$ processes of the type (\ref{eq:quantum_process}) for each
mode $i$. Evidently, the process (\ref{eq:quantum_process}) transports
information (which could be previously scrambled by a unitary evolution
of the black hole) encoded in the (base) qubits $\left|q_{i}\right\rangle $
($q_{i}=0,1$) from the black hole modes $a_{i}$ to the outgoing
radiation modes $c_{i}$. Moreover, the mode levels inside the black
hole ($a_{i}$ and $b_{i}$) become gradually emptied (classically,
the black hole gradually shrinks), and furthermore, the primarily
entangled initial state on the LHS of (\ref{eq:quantum_process})
becomes unentangled on the RHS.

\section{Unitarity}

To explicitly show unitarity of the qubit transformation (\ref{eq:quantum_process}),
it is sufficient to construct a corresponding unitary operator performing
the required transformation (\ref{eq:quantum_process}). To this end,
let us first define two auxiliary (parameter-dependent) orthonormal
bases in the tensor product Hilbert space $\mathcal{H}_{i}=\mathcal{H}_{a_{i}}\times\mathcal{H}_{b_{i}}\times\mathcal{H}_{c_{i}}$,
where we have introduced the three Hilbert spaces (of complex dimension
2, each) for all types of the involved modes: $a_{i}\in\mathcal{H}_{a_{i}}$,
$b_{i}\in\mathcal{H}_{b_{i}}$, $c_{i}\in\mathcal{H}_{c_{i}}$ ($\dim_{\mathbb{C}}\mathcal{H}_{i}=8$).
The total Hilbert space is then $\mathcal{H}=\bigotimes_{i=1}^{n}\mathcal{H}_{i}$.

The first (unprimed) orthonormal base parameterized by $\omega_{i}$
is $\left\{ \left|E_{\Lambda}\left(\omega_{i}\right)\right\rangle _{a_{i}b_{i}c_{i}}\right\} _{\Lambda=0}^{7}$,
and it consists of the following set of states/vectors:
\begin{equation}
\begin{aligned}\left|E_{0}\right\rangle _{a_{i}b_{i}c_{i}} & =\left|0\right\rangle _{a_{i}}\left(\cos\omega_{i}\left|0\right\rangle _{b_{i}}\left|0\right\rangle _{c_{i}}+\sin\omega_{i}\left|1\right\rangle _{b_{i}}\left|1\right\rangle _{c_{i}}\right)\\
\left|E_{1}\right\rangle _{a_{i}b_{i}c_{i}} & =\left|0\right\rangle _{a_{i}}\left|0\right\rangle _{b_{i}}\left|1\right\rangle _{c_{i}}\\
\left|E_{2}\right\rangle _{a_{i}b_{i}c_{i}} & =\left|0\right\rangle _{a_{i}}\left|1\right\rangle _{b_{i}}\left|0\right\rangle _{c_{i}}\\
\left|E_{3}\right\rangle _{a_{i}b_{i}c_{i}} & =\left|0\right\rangle _{a_{i}}\left(-\sin\omega_{i}\left|0\right\rangle _{b_{i}}\left|0\right\rangle _{c_{i}}+\cos\omega_{i}\left|1\right\rangle _{b_{i}}\left|1\right\rangle _{c_{i}}\right)\\
\left|E_{4}\right\rangle _{a_{i}b_{i}c_{i}} & =\left|1\right\rangle _{a_{i}}\left(\cos\omega_{i}\left|0\right\rangle _{b_{i}}\left|0\right\rangle _{c_{i}}+\sin\omega_{i}\left|1\right\rangle _{b_{i}}\left|1\right\rangle _{c_{i}}\right)\\
\left|E_{5}\right\rangle _{a_{i}b_{i}c_{i}} & =\left|1\right\rangle _{a_{i}}\left|0\right\rangle _{b}\left|1\right\rangle _{c_{i}}\\
\left|E_{6}\right\rangle _{a_{i}b_{i}c_{i}} & =\left|1\right\rangle _{a_{i}}\left|1\right\rangle _{b_{i}}\left|0\right\rangle _{c_{i}}\\
\left|E_{7}\right\rangle _{a_{i}b_{i}c_{i}} & =\left|1\right\rangle _{a_{i}}\left(-\sin\omega_{i}\left|0\right\rangle _{b_{i}}\left|0\right\rangle _{c_{i}}+\cos\omega_{i}\left|1\right\rangle _{b_{i}}\left|1\right\rangle _{c_{i}}\right).
\end{aligned}
\label{eq:first_base}
\end{equation}

The second (primed) orthonormal base parameterized by $\theta_{i}$
is $\left\{ \left|E_{\Lambda}'\left(\theta_{i}\right)\right\rangle {}_{a_{i}b_{i}c_{i}}\right\} _{\Lambda=0}^{7}$,
and it is defined as:
\begin{equation}
\begin{aligned}\left|E_{0}'\right\rangle _{a_{i}b_{i}c_{i}} & =\left|0\right\rangle _{a_{i}}\left|0\right\rangle _{b_{i}}\left|0\right\rangle _{c_{i}}\\
\left|E_{1}'\right\rangle _{a_{i}b_{i}c_{i}} & =\cos\frac{\theta_{i}}{2}\left|0\right\rangle _{a_{i}}\left|0\right\rangle _{b_{i}}\left|1\right\rangle _{c_{i}}-\sin\frac{\theta_{i}}{2}\left|1\right\rangle _{a_{i}}\left|0\right\rangle _{b_{i}}\left|0\right\rangle _{c_{i}}\\
\left|E_{2}'\right\rangle _{a_{i}b_{i}c_{i}} & =\left|0\right\rangle _{a_{i}}\left|1\right\rangle _{b_{i}}\left|0\right\rangle _{c_{i}}\\
\left|E_{3}'\right\rangle _{a_{i}b_{i}c_{i}} & =\left|0\right\rangle _{a_{i}}\left|1\right\rangle _{b_{i}}\left|1\right\rangle _{c_{i}}\\
\left|E_{4}'\right\rangle _{a_{i}b_{i}c_{i}} & =\sin\frac{\theta_{i}}{2}\left|0\right\rangle _{a_{i}}\left|0\right\rangle _{b_{i}}\left|1\right\rangle _{c_{i}}+\cos\frac{\theta_{i}}{2}\left|1\right\rangle _{a_{i}}\left|0\right\rangle _{b_{i}}\left|0\right\rangle _{c_{i}}\\
\left|E_{5}'\right\rangle _{a_{i}b_{i}c_{i}} & =\left|1\right\rangle _{a_{i}}\left|0\right\rangle _{b}\left|1\right\rangle _{c_{i}}\\
\left|E_{6}'\right\rangle _{a_{i}b_{i}c_{i}} & =\left|1\right\rangle _{a_{i}}\left|1\right\rangle _{b_{i}}\left|0\right\rangle _{c_{i}}\\
\left|E_{7}'\right\rangle _{a_{i}b_{i}c_{i}} & =\left|1\right\rangle _{a_{i}}\left|1\right\rangle _{b_{i}}\left|1\right\rangle _{c_{i}}.
\end{aligned}
\label{eq:second_base}
\end{equation}

The explicitly unitary transformation $U_{i}\left(\theta_{i}\right)$
can now be constructed from the two bases (\ref{eq:first_base}) and
(\ref{eq:second_base}) in a standard way as
\begin{equation}
U_{i}\left(\theta_{i}\right)=\sum_{\Lambda=0}^{7}\left|E_{\Lambda}'\left(\theta_{i}\right)\right\rangle _{a_{i}b_{i}c_{i}}\left\langle E_{\Lambda}\left(\pi^{-1}\theta_{i}\Omega_{i}\right)\right|_{a_{i}b_{i}c_{i}},\label{eq:unitary_transformation}
\end{equation}
where the auxiliary parameter (``time'') $\theta_{i}$ satisfies:
$0\leq\theta_{i}\leq\pi$ (cf.\ \citep{Osuga2018}). Evidently, for
$\theta_{i}=\pi$, the unitary operator (\ref{eq:unitary_transformation})
performs the required transformation (\ref{eq:quantum_process}).

Thus, finally, expressing the total unitary transformation as a tensor
product $U\left(\pi\right)=\bigotimes_{i=1}^{n}U_{i}\left(\pi\right)$,
we can write $U\left(\pi\right)\left|\Psi_{0}\right\rangle =\left|\Psi_{1}\right\rangle $,
where the final state $\left|\Psi_{1}\right\rangle $ assumes the
explicit form
\begin{equation}
\left|\Psi_{1}\right\rangle =\prod_{i=1}^{n}\left|0\right\rangle _{a_{i}}\prod_{i=1}^{n}\left|0\right\rangle _{b_{i}}\sum_{q_{1},q_{2},\ldots,q_{n}=0}^{1}A_{q_{1}q_{2}\cdots q_{n}}\prod_{i=1}^{n}\left|q_{i}\right\rangle _{c_{i}}.\label{eq:final_state}
\end{equation}
It follows from Eq.(\ref{eq:final_state}) that the whole black hole
information encoded in the amplitudes $A_{q_{1}q_{2}\cdots q_{n}}$
has been transferred from the black hole modes $a_{i}$ to the outgoing
radiation modes $c_{i}$, whereas the black hole modes $a_{i}$, $b_{i}$
are now in the vacuum state. Moreover, there is no entanglement between
modes inside and outside of the black hole.

One could wonder what happens if the final state $\left|\varPsi_{1}\right\rangle $
is ``time-evolved'' further. First of all, we should emphasize that
the proposed model is supposed to refer only to a prescribed period
of time. In the assumed units, it is $0\leq\theta\leq\pi$. Therefore,
at the end of the time evolution all $a_{i}$- and $b_{i}$-modes
are (see (\ref{eq:final_state})) in a vacuum state. But if we, purely
formally, time-evolved the final state further using Eq.(\ref{eq:unitary_transformation})
outside the domain of its supposed applicability, due to the presence
of trigonometric functions in (\ref{eq:first_base}) and (\ref{eq:second_base}),
we would observe (mode-dependent) oscillatory character of the evolution.
In particular, $a_{i}$- and $b_{i}$-states could get switched on
again.

\section{Final remarks}

The auxiliary parameter(s) $\theta_{i}$ can be used in (\ref{eq:unitary_transformation})
for two purposes. First of all, one can apply $\theta_{i}$ to gradually
switch-on the process (\ref{eq:quantum_process}); e.g., $\theta_{i}$
could be some decreasing functions of a curvature invariant as in
\citep{Osuga2018}. Alternatively, one could possibly try to extract
a corresponding ``Hamiltonian'' $\boldsymbol{H}$ (cf.\ Eq.(3.4)
in \citep{Osuga2018}) from the ``short-time'' limit: $U\left(\theta\right)=I-i\theta\boldsymbol{H}$.
However, the latter procedure is highly non-unique (see the following
paragraph).

We would like to draw the reader's attention to a minor technical
detail. Namely, the unitary operator (\ref{eq:unitary_transformation})
is introduced in a highly non-unique way, in the sense that there
is a large group of ($i$-dependent) unitary transformations (which
could be implemented, e.g., in terms of unitary transformations on
some subsets of vectors of the bases) changing the operator itself
but still performing the same transformation (\ref{eq:quantum_process})
(obviously, the same is true for the operator proposed by Osuga and
Page in \citep{Osuga2018}). This non-uniqueness follows from the
fact that there is large freedom in determining the unitary operator
\textemdash{} the freedom is only restricted by the condition/process
(\ref{eq:quantum_process}) which involves only some very particular
states. The action of the operator on other states is undetermined
by (\ref{eq:quantum_process}), and hence it is arbitrary.

Recapitulating, as a final remark, we would like to add that our idea
to use precisely the state (\ref{eq:initial_state}) as an initial
state (with fermionic Hawking states included), and (\ref{eq:final_state})
as a final state (with vacua for the black hole modes $a_{i}$ and
$b_{i}$) follows from the fact that it was our intention to have
a possibility to interpret qubits a little bit more realistically,
as possible fermionic particles, rather than as only purely formal
entities.

\bibliographystyle{unsrtnat}
\bibliography{unitary_toy_qubit_transport_model_for_black_hole_evaporation}

\begin{thebibliography}{16}
\providecommand{\natexlab}[1]{#1}
\providecommand{\url}[1]{\texttt{#1}}
\expandafter\ifx\csname urlstyle\endcsname\relax
  \providecommand{\doi}[1]{doi: #1}\else
  \providecommand{\doi}{doi: \begingroup \urlstyle{rm}\Url}\fi

\bibitem[Harlow(2016)]{Harlow2016a}
Daniel Harlow.
\newblock {Jerusalem Lectures on Black Holes and Quantum Information}.
\newblock \emph{Rev. Mod. Phys.}, 88:\penalty0 015002, 2016.
\newblock \doi{10.1103/RevModPhys.88.015002}.

\bibitem[Marolf(2017)]{Marolf2017a}
Donald Marolf.
\newblock {The Black Hole information problem: past, present, and future}.
\newblock \emph{Rept. Prog. Phys.}, 80\penalty0 (9):\penalty0 092001, 2017.
\newblock \doi{10.1088/1361-6633/aa77cc}.

\bibitem[Polchinski(2017)]{Polchinski2017c}
Joseph Polchinski.
\newblock {The Black Hole Information Problem}.
\newblock In \emph{{Proceedings, Theoretical Advanced Study Institute in
  Elementary Particle Physics: New Frontiers in Fields and Strings (TASI 2015):
  Boulder, CO, USA, June 1--26, 2015}}, pages 353--397, 2017.
\newblock \doi{10.1142/10270}.

\bibitem[Chakraborty and Lochan(2017)]{Chakraborty2017a}
Sumanta Chakraborty and Kinjalk Lochan.
\newblock {Black Holes: Eliminating Information or Illuminating New Physics?}
\newblock \emph{Universe}, 3\penalty0 (3):\penalty0 55, 2017.
\newblock \doi{10.3390/universe3030055}.

\bibitem[Unruh and Wald(2017)]{Unruh2017}
William~G. Unruh and Robert~M. Wald.
\newblock {Information Loss}.
\newblock \emph{Rept. Prog. Phys.}, 80\penalty0 (9):\penalty0 092002, 2017.
\newblock \doi{10.1088/1361-6633/aa778e}.

\bibitem[Giddings(2012{\natexlab{a}})]{Giddings2012}
Steven~B. Giddings.
\newblock {Black holes, quantum information, and unitary evolution}.
\newblock \emph{Phys. Rev.}, D85:\penalty0 124063, 2012{\natexlab{a}}.
\newblock \doi{10.1103/PhysRevD.85.124063}.

\bibitem[Giddings(2012{\natexlab{b}})]{Giddings2012a}
Steven~B. Giddings.
\newblock {Models for unitary black hole disintegration}.
\newblock \emph{Phys. Rev.}, D85:\penalty0 044038, 2012{\natexlab{b}}.
\newblock \doi{10.1103/PhysRevD.85.044038}.

\bibitem[Giddings and Shi(2013)]{Giddings2013d}
Steven~B. Giddings and Yinbo Shi.
\newblock {Quantum information transfer and models for black hole mechanics}.
\newblock \emph{Phys. Rev.}, D87\penalty0 (6):\penalty0 064031, 2013.
\newblock \doi{10.1103/PhysRevD.87.064031}.

\bibitem[Mathur(2009{\natexlab{a}})]{Mathur2009}
Samir~D. Mathur.
\newblock {The Information paradox: A Pedagogical introduction}.
\newblock \emph{Class. Quant. Grav.}, 26:\penalty0 224001, 2009{\natexlab{a}}.
\newblock \doi{10.1088/0264-9381/26/22/224001}.

\bibitem[Mathur(2009{\natexlab{b}})]{Mathur2009b}
Samir~D. Mathur.
\newblock {What Exactly is the Information Paradox?}
\newblock \emph{Lect. Notes Phys.}, 769:\penalty0 3--48, 2009{\natexlab{b}}.
\newblock \doi{10.1007/978-3-540-88460-6_1}.

\bibitem[Avery(2013)]{Avery2013}
Steven~G. Avery.
\newblock {Qubit Models of Black Hole Evaporation}.
\newblock \emph{JHEP}, 01:\penalty0 176, 2013.
\newblock \doi{10.1007/JHEP01(2013)176}.

\bibitem[Osuga and Page(2018)]{Osuga2018}
Kento Osuga and Don~N. Page.
\newblock {Qubit Transport Model for Unitary Black Hole Evaporation without
  Firewalls}.
\newblock \emph{Phys. Rev.}, D97\penalty0 (6):\penalty0 066023, 2018.
\newblock \doi{10.1103/PhysRevD.97.066023}.

\bibitem[Mathur and Plumberg(2011)]{Mathur2011}
Samir~D. Mathur and Christopher~J. Plumberg.
\newblock {Correlations in Hawking radiation and the infall problem}.
\newblock \emph{JHEP}, 09:\penalty0 093, 2011.
\newblock \doi{10.1007/JHEP09(2011)093}.

\bibitem[M{\"{u}}ck(2016)]{Mueck2016}
Wolfgang M{\"{u}}ck.
\newblock {Hawking radiation is corpuscular}.
\newblock \emph{Eur. Phys. J.}, C76\penalty0 (7):\penalty0 374, 2016.
\newblock \doi{10.1140/epjc/s10052-016-4233-3}.

\bibitem[Broda(2017)]{Broda2017}
Bogus{\l}aw Broda.
\newblock {Total spectral distributions from Hawking radiation}.
\newblock \emph{Eur. Phys. J.}, C77\penalty0 (11):\penalty0 756, 2017.
\newblock \doi{10.1140/epjc/s10052-017-5336-1}.

\bibitem[Almheiri et~al.(2013)Almheiri, Marolf, Polchinski, Stanford, and
  Sully]{Almheiri2013}
Ahmed Almheiri, Donald Marolf, Joseph Polchinski, Douglas Stanford, and James
  Sully.
\newblock {An Apologia for Firewalls}.
\newblock \emph{JHEP}, 09:\penalty0 018, 2013.
\newblock \doi{10.1007/JHEP09(2013)018}.

\end{thebibliography}

\end{document}